# Structural and vibrational study of Zn(IO$_3$)$_2$ combining high-pressure experiments and density-functional theory

A. Liang[1], C. Popescu[2], F. J. Manjon[3], P. Rodriguez-Hernandez[4], A. Muñoz[4], Z. Hebboul[5], and D. Errandonea[1]

[1]*Departamento de Física Aplicada-ICMUV-MALTA Consolider Team, Universitat de València, c/Dr. Moliner 50, 46100 Burjassot (Valencia), Spain*
[2]*CELLS-ALBA Synchrotron Light Facility, Cerdanyola del Vallès, 08290 Barcelona, Spain*
[3]*Instituto de Diseño para la Fabricación y Producción Automatizada, MALTA Consolider Team, Universitat Politècnica de València, Camí de Vera s/n, 46022 València, Spain*
[4]*Departamento de Física and Instituto de Materiales y Nanotecnología, MALTA Consolider Team, Universidad de La Laguna, 38206 La Laguna, Tenerife, Spain*
[5]*Laboratoire Physico-Chimie des Matériaux (LPCM), University Amar Telidji of Laghouat, BP 37G, Ghardaïa Road, Laghouat 03000, Algeria*

# ABSTRACT

We report a characterization of the high-pressure behavior of zinc-iodate, $Zn(IO_3)_2$. By the combination of x-ray diffraction, Raman spectroscopy, and first-principles calculations we have found evidence of two subtle isosymmetric structural phase transitions. We present arguments relating these transitions to a non-linear behavior of phonons and changes induced by pressure on the coordination sphere of the iodine atoms. This fact is explained as a consequence of the formation of metavalent bonding at high-pressure which is favored by the lone-electron pairs of iodine. In addition, the pressure dependence of unit-cell parameters, volume, and bond is reported. An equation of state to describe the pressure dependence of the volume is presented, indicating that $Zn(IO_3)_2$ is the most compressible iodate among those studied up to now. Finally, phonon frequencies are reported together with their symmetry assignment and pressure dependence.

# I. INTRODUCTION

Metal iodates form a group of materials that have been intensively studied as promising nonlinear optics (NLO) materials [1–3] in the visible, near, and mid infrared (IR) ranges of the electromagnetic spectrum. Moreover, recent density-functional theory (DFT) calculations reveal that metal iodates exhibit large or even giant elastocaloric (EC) effect, which made them promising materials for developing solid state cooling technologies [4,5]. These materials are of interest also for fundamental research owing to the presence of stereochemically active lone electron pairs (LEPs) on iodine atoms.

Of special interest are the studies of the crystal structure and vibration behavior of metal iodates at high pressure (HP). Recently, we have reported a comprehensive crystal structure and vibration study of $Fe(IO_3)_3$ under compression employing X-ray diffraction, IR spectroscopy, Raman scattering (RS) measurements, and first-principles DFT calculations [6,7]. Three isostructural phase transitions (IPTs) were found at 1.5, 5.7, and 22 GPa, respectively. The first two transitions do not involve detectable discontinuities in the unit-cell volume, but the third transition has associated a large volume collapse, being a first-order transition. The sequence of transformations is accompanied by a gradual increase of iodine (I) coordination from three to six [6]. The rich phase transition sequence and coordination change is mostly due to changes associated with the LEPs of I atoms [7]. As we have demonstrated in the previous HP-study of $Fe(IO_3)_3$, iodates seem to exhibit many unusual and interesting properties and changes at HP [6,7]. However, until now, except for $Fe(IO_3)_3$, only the behavior of the crystal structure of $LiIO_3$ [8–10], $KIO_3$ [11] and $AgIO_3$ [12] has been studied at HP. This does not include other properties, such as lattice vibrations, electronic density of states, and electronic band structure. For $Zn(IO_3)_2$, its crystal structure [13–15], Raman and IR spectrum [16,17] have been well studied at ambient conditions, but research at HP is not available.

Here we extend the HP-study on another metal iodate, $Zn(IO_3)_2$. At ambient conditions, $Zn(IO_3)_2$ crystallizes in a monoclinic structure (space group: $P2_1$, Z=4) [15], where $ZnO_6$ octahedral units are bridged by $IO_3$ triangular pyramids (see Fig. 1). In the $IO_3^-$ anion, iodine, with a valence configuration $5s^25p^5$, is pentavalent and forms three covalent bonds with oxygen by sharing its p electrons, leaving the 5s electrons free to orient along the c-axis, showing an $IO_3E$ configuration, being E the LEP. This characteristic of the $IO_3^-$ anion triggers an interesting and unusual behavior in $Fe(IO_3)_3$ at HP [6,7,11,12]. Apart from that, the LEP-induced anharmonicity can provide an efficient way to minimize the lattice thermal conductivity [18,19]. These characteristics are shared by the crystal structures of $Zn(IO_3)_2$ and $Fe(IO_3)_3$, which suggests that $Zn(IO_3)_3$ could potentially have a quite interesting EC effect at HP.

In this paper, HP X-ray diffraction (HP-XRD), HP Raman scattering (HP-RS), and DFT calculations have been used to investigate the HP behavior of $Zn(IO_3)_2$. Two symmetry-preserving phase transitions have been found. Their existence has been related to changes found in the iodine atom coordination. We have also observed the existence of soft-phonon modes, which have been found in the high-frequency range in HP-RS studies. These vibrations have been associated with the vibration behavior of I-O bonds. All of the experimental data will be compared with DFT simulations, giving a deep insight into the structural and vibrational behavior of $Zn(IO_3)_2$ under pressure.

## II. METHODS

### A. Sample preparation

For the study of $Zn(IO_3)_2$, micron-sized needle-like crystals were synthesized from an aqueous solution [20]. $Zn(IO_3)_2$ was prepared by mixture of potassium iodate $KIO_3$ (2 mmol, 99.5% purity) and zinc chloride $ZnCl_2$ (1 mmol, 98% purity) in de-ionized water: $2KIO_3 + ZnCl_2 \rightarrow Zn(IO_3)_2 + 2KCl$. The reagents were purchased from Fluka ($KIO_3$) and Riedel-deHeän ($ZnCl_2$). The reaction mixtures were slowly evaporated and maintained at 60° C for four days, leading to the formation of white precipitations. We filtrate and washing the crystals with de-ionized water.

To confirm the purity and crystal structure of $Zn(IO_3)_2$, we performed powder XRD measurements at room conditions (see Fig. 2) using a XPERT Pro diffractometer from Panalytical in a reflection Bragg-Brentano geometry and employing monochromatic Cu $K_{\alpha 1}$ radiation. The Rietveld refinement method [21] was adopted to analyze the phase and lattice parameter information. The structural information of $Zn(IO_3)_2$ reported by Liang *et al.* [15] (space group $P2_1$) was used as a starting point to do the refinement. We have found in the refinement that all the peak positions match very well and that the intensity of most peaks agrees with those already reported except for some peaks between 30º-50º due to the partially preferred direction of the needle-shaped sample. The lattice parameters obtained from our refinements and DFT calculations are listed in Table 1 and compared with the values reported in the literature [15]. The agreement regarding unit-cell parameters is good. In fact, calculations, agree better with present experiments than with previous ones.

## B. High-pressure experiment details

Angle-dispersive powder HP-XRD experiments were performed at room temperature (RT) up to 27.8 GPa employing a membrane diamond-anvil cell (DAC), with diamond culets of 400 μm. For these experiments the $Zn(IO_3)_2$ sample was grinded to obtain a fine powder which was loaded in a 180-μm diameter hole drilled on a stainless-steel gasket pre-indented to a thickness of 40 μm. Pressure was measured using the equation of state of gold [22] and a 16:3:1 methanol:ethanol-water (MEW) mixture was used as the pressure-transmitting medium. Experiments were performed at the BL04-MSPD of ALBA synchrotron [23] with a monochromatic X-ray beam ($\lambda$ = 0.4246 Å) focused to a 20 μm × 20 μm full with at half maximum spot.

HP-RS measurements up to 12 GPa were performed in $Zn(IO_3)_2$ also with a DAC, where sample and ruby chips were loaded in the 150-μm diameter hole of a pre-indented steel gasket. A 4:1 methanol-ethanol mixture was used as pressure-transmitting medium [24,25]. Pressure was determined using the ruby fluorescence scale [26]. We paid special attention during sample loading to avoid the hydration of the metal iodates [27]. We also made a careful loading avoiding sample bridging between diamond anvils [28]. Measurements were performed in a backscattering geometry employing a HeNe laser (632.8 nm) and a Horiba Jobin Yvon LabRAM HR UV microspectrometer with a spectral resolution better than 2 $cm^{-1}$.

## C. Ab-initio simulations

Calculations of the total energy as a function of pressure for $Zn(IO_3)_2$ were performed using the Vienna ab initio simulations package (VASP) [29–32] within the framework of the density functional theory (DFT) [33] and the projector-augmented wave pseudopotential method (PAW) [34,35]. A high plane-wave energy cutoff of 540 eV was used to obtain accurate results. The exchange-correlation energy was described within the generalized gradient approximation (GGA) with the PBE for solids prescription (PBEsol) [36]. The Monkhorst-Pack scheme [37] was employed to discretize the Brillouin zone (BZ) integrations with a 4 x 8 x 4 mesh to ensure accuracy. We tested that this mesh is sufficient to avoid all non-systematic discretization errors

With this procedure, very high convergences of 1 meV per formula unit in the total energy are achieved and in the optimized configurations, the forces are lower than 2meV/Å per atom and the deviation of the stress tensor from a diagonal hydrostatic form is less than 0.1 GPa. All the structural parameters were obtained at selected optimized volumes. The agreement of the calculations with experiments regarding the crystal structure (within 1%, see Table 1), supports that our DFT calculations describe properly $Zn(IO_3)_2$.

After optimizing the crystal structure at different pressures a study of the phonons was performed at the zone center (Γ point) of the BZ using the direct method [38]. These calculations also allow identifying the symmetry and eigenvectors of the phonon modes at the Γ point.

## III. RESULTS AND DISCUSSION

### A. Structural study

Fig. 3 shows HP-XRD patterns of $Zn(IO_3)_2$ at selected pressures. At the lowest pressure, all of the peaks can be assigned to the monoclinic structure we described above. Some peaks, like (120) and (041) peaks, show a relative intensity change with increasing pressure. This is due to preferred orientations related to the needle-shaped sample. We must note that preferred orientations were present at ambient pressure and become more obvious as pressure increases. From ambient to the highest pressure, all the XRD patterns can be identified in the same space group. However, as we will discuss below we will show that subtle changes observed in the pressure evolution of lattice parameters suggest the existence of isosymmetric structural phase transitions. The changes induced by pressure in XRD patterns are reversible as can be seen in Fig. 3 in the XRD pattern measured at ambient pressure after the decompression process.

Under compression, most diffraction peaks shift to higher angles, as expected [25,39]. However, some peaks, like (100), (120), and (3-41) peaks, shift to lower angles under compression. After a deep inspection of the d-spacing evolution under pressure (see Fig. S1, Supplementary Information [40]), we have found that the corresponding interplanar distance (d-spacing) of the above mentioned peaks first increases with pressure and then remain almost unchanged or reduce to a very small extent. Moreover, it must be noticed that d-spacings of (hk0) peaks change much less than the (hkl) peaks with l≠0. Lattice planes, like (001), (1-11), (101), (002), and (102) have a high-pressure dependence comparable to other lattice planes. Therefore, we can ascribe the different shift rates of the XRD peaks, and even the opposite shifts of some peaks, to the anisotropic compressibility of $Zn(IO_3)_2$ which we will discuss in detail later. Another consequence of the anisotropic compression of $Zn(IO_3)_2$ is the merging and crossing of several XRD peaks (or d-spacings).

Fig. 4 shows the experimental and theoretical pressure dependence of the unit-cell volume of $Zn(IO_3)_2$. A continuous change of the experimental unit-cell volume of the monoclinic structure is found along the whole pressure range up to 28 GPa with no appreciable jump in volume at any pressure. A 3$^{rd}$ order Birch-Murnaghan (BM) [41,42] equation of state (EOS) was adopted to fit the volume within the whole pressure range. The bulk modulus is 21.6 (0.7) and 18.4 (1.5) GPa according to experiments and calculations and the corresponding pressure derivate is 7.0 (0.3) and 6.7 (0.7), respectively. Experiment and calculations describe a similar high compressibility behavior of $Zn(IO_3)_2$ at low pressures and the minor difference between them is within a reasonable error [43]. The difference becomes more pronounced as the pressure is increased, in opposition to the general systematic followed by oxides [43]. This could be a consequence of the formation of new I-O bonds under HP (as we will discuss later) which enhances localization of electronic states (in opposition to the electro delocalization induced by pressure in oxides). In this context, we must note that while our experimental data can be reasonably fitted to a single 3$^{rd}$ order EOS, our theoretical data do not fit nicely to a 3$^{rd}$ order EOS along the whole pressure range, especially between 3 and 10 GPa. This fact could be related to the occurrence of isosymmetric phase transitions in $Zn(IO_3)_2$, which would be discussed in the following paragraphs. A, EOS of the theoretical results in the 0-3 GPa range provides a bulk modulus of 18(2) GPa that agrees better with our experimental results.

In our previous study [7], we summarized the bulk modulus of some metal iodates, including $Fe(IO_3)_3$ [7], $LiIO_3$ [9,10], $KIO_3$ [11], and $LiZn(IO_3)_3$ [44]. The bulk modulus of $Zn(IO_3)_2$ is equivalent to that of the ambient-pressure phase of $KIO_3$ [11], about two thirds of that of $LiIO_3$ reported by our calculations [7] and Hu et al [9], and two fifths of that of the low-pressure phase of $Fe(IO_3)_3$ [7] and $LiZn(IO_3)_3$ [44]. Thus $Zn(IO_3)_2$ is one of the most compressible iodates, being as compressible as metal-organic frameworks [45], rare-gas solids [46], and group-XV oxides containing cationic LEPs, such as $As_2O_3$ [47–50], $Sb_2O_3$ [51,52], and $Bi_2O_3$ [53,54].

In order to further characterize volume changes, we have used a 3$^{rd}$ order BM EOS to describe the pressure dependence of the $ZnO_6$, $I_1O_6$, and $I_2O_6$ octahedral volumes (here we assume a 3+3 coordination of I). In this scenario, we obtain zero-pressure bulk moduli of 66(2), 26(2) and 15(1.3) GPa, with corresponding pressure derivatives of 9.2 (0.5), 5.3 (0.5) and 5.3 (0.5), respectively. Since we already know that the bulk modulus of bulk $Zn(IO_3)_2$ is 21.6 GPa in the experiment and 18.4 GPa in the calculations, it is clear that the bulk modulus of the crystal is mainly determined by the compression of $IO_6$ units, rather than the compression of the $ZnO_6$ unit. This result is similar to what has been found in $Fe(IO_3)_3$ [7].

From Rietveld refinement of HP-XRD patterns, we have studied the evolution of unit-cell parameters of $Zn(IO_3)_2$ under pressure and compared them with the results of DFT calculations (see Fig. 5). There is an excellent agreement between experimental and theoretical results. Thus, we conclude that our calculations describe properly the crystal behavior under compression. As can be seen in Fig. 5, the behavior is strongly anisotropic. In addition, distinctive behaviors are observed for pressure ranges separated by vertical lines in Fig. 5 (P < 3.4 GPa, 3.4 GPa < P < 8.9 GPa, P > 8.9 GPa). In the low-pressure range, the linear compressibilities of a- and b-axis are similar and very small. In the medium-pressure range, they are even negative. Finally, in the high-pressure range, both unit-cell parameters show a similar slow decrease with pressure. In contrast, the behavior of the c-axis is very different, being much more compressible than the other two axes, at least in the two first pressure ranges, and showing a strong decrease of the compressibility in the third pressure range. The much higher compressibility of the c-axis at low pressures is a consequence of the quasi-layered structure formed by $IO_3$ tetrahedron and connected by $ZnO_6$ octahedron along this direction (see Fig. 1b). Note that the structure is not strictly layered because of $ZnO_6$ octahedra connect alternate $IO_3$ layers; however, $IO_3E$ units lead to considerable empty spaces in the structure along the c-axis where the iodine LEPs are located. The collapse of the space between $IO_3$ layers contributes to the high compressibility of the c-axis and favors the formation of three additional I-O bonds. In contrast, the alternate arrangement of the $ZnO_6$ octahedron and $IO_3$ tetrahedron along a- and b-axis strengthen the force and resist compression along those directions.

To highlight the different behavior of the axes in $Zn(IO_3)_2$, we plot in Fig. S2 the ratio between lattice parameters as a function of pressure. It can be seen that the b/a ratio is almost insensitive to pressure, but b/c and a/c have slope changes at 3.4 and 8.9 GPa that agree with those observed in the lattice parameters and the monoclinic gamma angle, as commented below.

We will comment now on the pressure dependence of the monoclinic γ angle (inset of Fig. 5). Before doing that, we would like to mention that at the lowest pressure (0.15 GPa) we got an experimental γ angle of 120.5(1)°, which is slightly higher than the ambient pressure value (120.37(8)°) but comparable within error bars. Upon compression, we have found both an experimental and theoretical S-like nonlinear pressure dependence of the monoclinic angle γ that is related to the symmetry increase of $ZnO_6$ and $IO_6$ polyhedra under pressure, with changes of slope near 3.4 and 8.9 GPa, in good agreement with the changes observed in the lattice parameters.

Based upon the good agreement between calculations and experiments regarding the pressure dependence of structural parameters and the difficulty to obtain reliable atomic positions from our HP-XRD measurements due to the influence of preferred orientations, we have performed a more detailed analysis of the theoretical data to investigate the behavior of the monoclinic structure near 3.4 and 8.9 GPa. In particular, we have used those data to calculate the pressure dependence of the theoretical distortion index of both polyhedra as calculated by using VESTA [55] (see Fig. S3 in the Supplementary Material [40]). The decrease of the distortion index of both polyhedra upon compression suggests a symmetry increase of both polyhedra under pressure. An almost linear decrease of the distortion index of $IO_6$ units is observed. Additionally, an S-like nonlinear decrease of the distortion index of $ZnO_6$ units with changes in slopes near 3.8 and 8.9 GPa are observed. The sum of the pressure effect on both polyhedra cause the nonlinear change of γ under pressure. Moreover, the changes observed in both polyhedra lead to: i) a symmetrization of the $ZnO_6$ octahedron, and ii) the gradual formation of $IO_6$ octahedra; i.e. an increase of I coordination due to the rapid decrease of the second-nearest neighbor I-O distances (represented in dashed lines in Fig. 1). These are evidences of the anisotropic compression of $Zn(IO_3)_2$. We will comment more on this unusual behavior below when taking about the behavior of Zn-O and I-O distances.

We think the structural changes observed at 3.4 and 8.9 GPa are caused by two subtle isostructural phase transitions (IPTs) [56,57]. More evidence of the existence of the two IPTs is shown in Figs. S4-S12 in the Supplementary Material [40], where the pressure dependence of the theoretical atomic coordinates of Zn, I, and O atoms are presented. Just as an example, we will comment on the calculated y atomic coordinate of Zn (see Fig. S4). Its initial value 0.7460, decreases sharply to 0.7452 at 2.5 GPa, and then increases to 0.7482 at 8.9 GPa, above this pressure, there is a sudden slope change, increasing slowly to 0.7488 at 21.9 GPa. The whole picture of the evolution is non-linear having an S-like shape. A similar S-like evolution of the atomic positions can be seen also in other atoms, thus supporting the two IPTs already commented. Moreover, the lack of a volume jump near these two pressures (see Fig. 4) suggest that the two IPTs are not of first-order character.

More evidence for the two IPTs can be found in the theoretical pressure dependence of the Zn-O and I-O distances and, more importantly, in the experimental and calculated Raman-mode evolution under pressure to be discussed in the next section. Regarding the pressure dependence of the Zn-O distances (see Fig. 6), one can notice an unusual S-like behavior for most Zn-O bond-distances, except for Zn-$O_3$. The behavior of the three short Zn-O bonds is consistent with the proposed IPTs at 3.4 and 8.9 GPa, with a decrease of the bond distances up to 3.8 GPa, an increase above this pressure and up to 8.9 GPa followed by a new decrease above this pressure. We can observe a considerable decrease of the three longest Zn-O bond distances to achieve an almost regular $ZnO_6$ polyhedron above 9.8 GPa, as already pointed out by the polyhedral distortion (Fig. S3).

As regards the pressure dependence of the I-O distances, we must recall that there are two different atomic positions for iodine, $I_1$, and $I_2$. $I_1$ is the iodine in the bottom and top of the unit cell when viewed from a-axis (see Fig. 1), while $I_2$ is in the middle. Here we will focus on the change of $I_1$-O bond distances with pressure (Fig. 7) because both I atoms show the same pressure behavior for their coordination polyhedron. As observed, there are three short I-O distances ($I_1$-$O_1$, $I_1$-$O_2$ and $I_1$-$O_3$) and three large distances above 2.5 Å at room pressure ($I_1$-$O'_1$, $I_1$-$O'_3$ and $I_1$-$O_4$, where $O'_1$ and $O'_3$ are second-neighbor $O_1$ and $O_3$ atoms). In the $IO_6$ polyhedron, these three second neighbor O atoms show distances above 2.5 Å and are not forming a bond with I at low pressure. As pressure increases these long distances show a strong decrease and gradually form new bonds. We can found a gradual I coordination change if we chose 2.48 Å as the maximum bonding distance [58]. In this way, with increasing pressure, $O'_1$, and $O_4$, which is on the side of the iodine LEPs, approaches I, first forming new I-O bonds with iodine at around 2.5 GPa, and then the I atom becomes six-fold coordinated above around 8 GPa due to the decrease of the I-$O'_3$ distance. On the other hand, at ambient pressure, the $ZnO_6$ octahedron is more symmetrical than the $IO_6$ octahedron, and compared with the change in the bond distance between Zn and O under pressure (the largest change is the change in the bond distance between the Zn and $O_6$ atoms, from 2.16 to 2.04 Å, 5.6% reduction), the change of the bond distance between I and O is in a larger range, the biggest change is the bond distance change between I and $O_3$ atom, shortened from 2.73 to 2.28 Å, a decrease of 16.5%. The great compressibility of the three longest $I_1$-O bond distances compared to the Zn-O bonds and the three shortest I-O bonds (Fig. 7), make the $IO_6$ octahedron to have a much larger compressibility than the $ZnO_6$ octahedron, as already commented.

The strong distortion of the $IO_6$ polyhedra at low pressures is due to the large stereoactivity of the iodine LEP. This stereoactivity decreases in strength with pressure and ultimately disappears in order to stablish the three new bonds above around 8 GPa, thus leading to a more regular $IO_6$ polyhedron as already commented. The pressure-induced reduction of the cation LEP stereoactivity can be found in some other materials, like $Fe(IO_3)_3$ [6,7], α-$Sb_2O_3$ [50,59,60], β-$Bi_2O_3$ [53,61], isostructural $Sb_2S_3$, $Sb_2Se_3$, and $Bi_2S_3$ [62], α-$As_2Te_3$ [63], $SbPO_4$ [64], and $As_2S_3$ [65]. On the other hand, we must say that the $I_1$-$O_1$ bond distance clearly shows a different pressure dependence before and after 3.4 GPa, thus supporting also the existence of an IPT around that pressure (see inset of Fig. 7).

Finally, we must note that there is an increase of the three shortest I-O distances that is in contrast to the decrease in the three largest I-O distances. The increase of the three shortest I-O distances will lead to the soft behavior of some Raman-active modes at HP that we will show in the next section. In this context, it must be stressed that an increase of the three shorter Zn-O bond distances with increasing pressure have also been observed between 3.8 and 8.9 GPa. The S-like behavior of the Zn-O bond distances with pressure is not normal. In particular, the increment of the cation-anion bond distance and the equalization of the different bond distances has been recently observed for As-S bonds in orpiment ($As_2S_3$) under compression [65]. In that case, it was ascribed to the formation of a new type of bond, named metavalent bonding, and ascribed to the sharing of electrons of two covalent As-S bonds (with 2 electrons per bond) in order to form four metavalent bonds (with 1 electron per bond). Therefore, we tentatively ascribe the S-like behavior of Zn-O bond distances to the competition of two effects that should be investigated in further works.

## B. Raman study

First, we will start discussing the Raman spectra of $Zn(IO_3)_2$ at ambient pressure (see Fig. 8). The RS spectrum at ambient pressure is very similar to those reported in the literature [16,17]. It can also be mentioned that the Raman spectrum of $Zn(IO_3)_2$ bears some resemblance to that of $Fe(IO_3)_3$ [6], despite the different crystal structures of the two compounds. In particular, all of them have Raman-active modes distributed in two isolated regions separated by a phonon gap. One high-frequency region for wavenumbers larger than 700 cm$^{-1}$, and other one for wavenumbers smaller than 450 cm$^{-1}$. In fact, the lowest region can be subdivided into two regions, a low-frequency region below 200 cm$^{-1}$ and a medium-frequency region between 200 and 450 cm$^{-1}$. The strongest mode is always in the high-frequency region. In this paper, we will show that the high-frequency modes can be linked to internal I-O vibrations inside the $IO_6$ polyhedra. In particular, the strongest mode is related to a symmetric stretching I-O vibration.

According to group theory, $Zn(IO_3)_2$ should have fifty-four vibrational modes with three acoustic modes (A+ 2B) and fifty-one Raman-active and IR modes (26A + 25B). We have observed twenty-two Raman-active modes out of fifty-one. In previous studies, twenty-one and sixteen modes have been reported [16,17]. Our results and those of previous works show a nice agreement (see Table 2). The calculated frequencies for all these modes together with a tentative mode assignment are also provided in Table 2. We must note that the experimental mode assignment in Table 2 is based on comparing the experimental frequencies at ambient pressure and their pressure coefficients at different pressure ranges with those obtained from DFT calculations. The agreement between experiments and calculations is similar to that found in $Fe(IO_3)_3$ [6]. The strongest Raman modes in $Zn(IO_3)_2$ and $Fe(IO_3)_3$ are at 782 and 790 cm$^{-1}$, respectively [6]. Indeed, this mode has been attributed to an A mode corresponding to the symmetric I-O stretching mode inside $IO_3$ units. Since I-O distances are very similar in both iodates, it is expected that both modes should have a similar frequency.

Now we focus on the HP behavior of the RS spectra in $Zn(IO_3)_2$ to continue our discussion (see Fig. 9). A change in the intensity of some modes is observed as pressure increases. For instance, the peak around 200 cm$^{-1}$ at 0.7 GPa is barely followed above 9 GPa. In addition, some peaks merge into one broad peak, like what happened to the soft mode of the high-frequency region around 720 cm$^{-1}$ at 0.7 GPa. In addition, we can observe two features in the RS spectra of $Zn(IO_3)_2$ under pressure that are similar to those reported in $Fe(IO_3)_3$ [6]: i) A reversible change of the HP-RS spectra, and b) a closing of the phonon gap between the medium- and high-frequency regions.

The pressure evolution of the Raman mode frequencies and the closing phonon gap can be better seen in Fig. 10. In the medium- and low-frequency regions, there is a close agreement between experimental and theoretical results. Under compression, there are some crossings and anti-crossings of modes but all vibrational modes correspond to a monoclinic crystal structure described by space group $P2_1$ up to 12 GPa. In the high-frequency region (Fig. 10 (c)), the behavior of the experimental modes can be described closely by calculations if we consider that theoretical calculations underestimate the experimental frequencies by ca. 70 cm$^{-1}$. All the phonons were fitted by a quadratic equation using the data below 3 GPa due to the S-like behavior of the mode evolution curve and the zero-pressure coefficients and Grüneisen parameters can be found in Table 2. The non-linear S-like curve, which can be found in the evolution of almost all the experimental and calculated modes, is a result of the two symmetry-preserving IPTs, the vertical dash green line indicated the phase transition pressure in HP-RS measurements, which is in excellent agreement with what we found in HP-XRD measurements. The non-linear pressure dependence of the Raman-active mode frequencies, related to IPTs, can also be found in $Fe(IO_3)_3$ and $\beta$-$Bi_2O_3$ [6,53,61].

Now we focus on the high-frequency region of the Raman spectrum (see Fig. 10(c)). First, we would like to mention that the experimental mode of highest frequency is a second-order Raman mode that is likely a combination of modes observed near 400-450 cm$^{-1}$ (all of them showing positive pressure coefficient). Second, we would like to comment on the observed phonon softening upon compression of many high-frequency modes. This is a common feature of all the calculated phonons and most experimental phonons of $Zn(IO_3)_2$. Based on our DFT calculation, we are capable of seeing the related atomic movements of all phonons by using J-ICE [66] and in particular, those corresponding to the high-frequency soft phonons (see Fig. S13 in the Supplementary material [40]). The soft phonons in our experimental data are mostly related to the stretch of the short I-O bonds, and the cause of the soft behavior of those phonons is the decrease of the force of these bonds as pressure increases due to the increase of the short I-O distances as pressure increases (see Fig. 7) as suggested by our DFT calculations.

To verify the hypothesis of the softening of the high-frequency phonons due to the increase of the short I-O bonds, we have assumed under a harmonic approximation that the force constant, k, of the I-O bond is a function of negative cubic of the average I-O bond distance, $(d_{I-O})^{-3}$. Since the frequency of the stretching mode, ω, can be written as ω = $(k/\mu)^{1/2}$, where μ is the mass, if the assumption is right and the mode softening is caused by the increase of the bond distance, then we can found a linear relationship between $\omega^{-2/3}$ and the bond distance. By using the calculated data, we plotted the theoretical frequencies of the soft modes with the average I-O bond distance in $IO_3$ units (see Fig. 11). We do see a clear linear relationship between the frequency and average I-O bond distance, thus the occurrence of the soft mode at the high-frequency region is induced by the increase of the three shortest I-O bond distances, which, as we explained before, is needed to accommodate the additional O atoms around iodine as pressure increases. The same HP behavior of the $IO_3$ units has been found in $Fe(IO_3)_3$ so it seems to be a common feature of compounds with iodine LEPs. We have to note that the softening of phonons related to the increase of short bonds with pressure has been also recently found in mineral orpiment ($As_2S_3$) and explained in the context of the formation of metavalent bonding at HP in compounds with stereoactive LEPs [65]. To explore this hypothesis in $Zn(IO_3)_2$ new studies are needed.

## IV. CONCLUSIONS

HP-XRD and HP-RS measurements were used to study the HP behavior of $Zn(IO_3)_2$ along with DFT theoretical calculations. The monoclinic crystal structure shows a high anisotropy and the most compressible axis is the c-axis. With a bulk modulus $B_0$ = 21.6 GPa, $Zn(IO_3)_2$ is the most compressible one among the whole reported metal iodine family. The compressibility of $Zn(IO_3)_2$ is governed by the compressibility of the $IO_6$ units. The symmetry of the experimental Raman-active modes was tentatively assigned and the softened phonons found at the high-frequency region were related to the increase of the shortest I-O bonds under pressure.

Two reversible symmetry-preserving IPTs were found and located around 2.5-3.4 and 8.9 GPa, without change of the space-group symmetry or crystal volume collapse. The two IPTs are associated with a gradual change from 3-fold to 6-fold of the I coordination under pressure, despite the most notable changes at the two pressure ranges are found for the y atomic coordinate of the Zn atom. The presence of the two IPTs leads to a nonlinear S-like pressure dependence of the unit-cell lattice parameters and Raman-active modes upon compression in both experiments and theoretical calculations, as well as in the calculated Zn-O and I-O bond distances. The gradual increase of I coordination with pressure due to the decrease of the large I-O bonds followed by an increase of the short I-O distances could be indicative of the formation of metavalent bonding at HP in compounds with stereoactive LEPs, as recently suggested for orpiment. To explore this hypothesis further studies are needed.

# ACKNOWLEDGMENTS

This work was supported by the Spanish Ministry of Science, Innovation and Universities under grant PID2019-106383GB-C41/42/43 and RED2018-102612-T (MALTA Consolider-Team Network) and by Generalitat Valenciana under grant Prometeo/2018/123 (EFIMAT). A. L. and D. E. would like to thank the Generalitat Valenciana for the Ph.D. fellowship GRISOLIAP/2019/025). C. P. is thankful for the financial support of the Spanish Mineco Project FIS2017-83295-P. Powder x-ray diffraction experiments were performed at the Materials Science and Powder Diffraction beamline of ALBA Synchrotron (Alba experiment 2019083663). Z. H. sincerely thanks to Professor I. K. Lefkaier - Laboratoire Physique des Matériaux (LPM) University of Laghouat Algeria - for the support and fruitful discussions.

## Tables

TABLE I. Summary of the lattice parameters and volume, $V_0$, of $Zn(IO_3)_2$ at ambient pressure from calculations (DFT) and experiments (Exp.). Previous experimental data from Ref. [15] are also included for comparison.

|  | DFT [a] | Exp.[a] | Exp. [15] |
|---|---|---|---|
| a (Å) | 5.469 | 5.465(4) | 5.469 |
| b (Å) | 10.832 | 10.952(8) | 10.938 |
| c (Å) | 5.054 | 5.129(4) | 5.1158 |
| $\gamma$ (°) | 120.514 | 120.37(8) | 120.000 |
| $V_0$ (Å$^3$) | 258.00 | 264.8(4) | 265.03 |

[a] This work,

TABLE II. Calculated and measured zero-pressure frequencies (ω), pressure coefficients and Grüneisen parameters of the Raman-active modes of $Zn(IO_3)_2$. Bulk moduli used to calculate Grüneisen parameters were obtained from our DFT calculated and HP-XRD data. Results are compared with previous experiments [16,17].

| mode | Theory ($B_0$ = 18.4 GPa) ω (cm$^{-1}$) | dω/dP (cm$^{-1}$/GPa) | γ | Experiment ($B_0$ = 21.6 GPa) ω (cm$^{-1}$) | dω/dP (cm$^{-1}$/GPa) | γ | ω [17] | ω [16] |
|---|---|---|---|---|---|---|---|---|
| A | 65.11 | 6.53 | 1.84 | 62 | 6.62 | 2.31 | 61 | |
| B | 68.35 | 1.06 | 0.28 | 67 | | | | |
| A | 79.19 | 3.66 | 0.85 | 80 | 3.63 | 0.99 | 80 | 80 |
| B | 82.56 | 2.49 | 0.55 | | | | | |
| A | 101.30 | 2.07 | 0.38 | | | | | |
| B | 110.08 | 2.30 | 0.38 | 101 | 2.85 | 0.61 | 100 | |
| A | 117.58 | 11.84 | 1.85 | 111 | 8.60 | 1.67 | 111 | 113 |
| B | 130.82 | 6.62 | 0.93 | | | | | |
| A | 136.49 | 3.64 | 0.49 | 128 | 11.73 | 1.99 | 130 | 132 |
| A | 142.00 | 5.96 | 0.77 | 149 | | | 148 | |
| B | 148.24 | 2.43 | 0.30 | | | | | |
| A | 155.14 | 6.44 | 0.76 | | | | | |
| B | 165.41 | 8.50 | 0.95 | 156 | 3.71 | 0.51 | 155 | 152 |
| A | 171.85 | 6.49 | 0.70 | 173 | 7.28 | 0.91 | 173 | 173 |
| B | 173.82 | 5.64 | 0.60 | | | | | |
| A | 177.56 | 8.20 | 0.85 | | | | | |
| B | 178.19 | 6.34 | 0.65 | | | | | |
| A | 185.36 | 3.07 | 0.30 | | | | | |
| B | 193.47 | 6.39 | 0.61 | 187 | 7.54 | 0.87 | 187 | 189 |
| A | 196.60 | 10.48 | 0.98 | | | | | |
| B | 212.11 | 3.76 | 0.33 | | | | | |
| A | 212.35 | 0.96 | 0.08 | | | | | |
| B | 224.42 | 9.35 | 0.77 | | | | | |
| B | 237.16 | 9.88 | 0.77 | | | | | |
| A | 239.73 | 7.41 | 0.57 | | | | | |
| B | 243.07 | 8.52 | 0.64 | | | | | |
| A | 262.65 | 5.34 | 0.37 | 264 | 5.77 | 0.47 | 265 | 267 |
| A | 307.48 | 3.41 | 0.20 | | | | | |
| B | 309.88 | 4.45 | 0.26 | | | | | |
| B | 325.39 | 1.54 | 0.09 | 327 | 3.80 | 0.25 | 327 | 327 |
| A | 332.93 | 5.28 | 0.29 | | | | | |
| A | 337.63 | 4.90 | 0.27 | | | | | |
| B | 344.67 | 5.82 | 0.31 | 351 | 5.77 | 0.35 | 351 | 354 |

| | | | | | | | |
|---|---|---|---|---|---|---|---|
| A | 374.06 | 5.22 | 0.26 | | | | |
| B | 375.29 | 3.13 | 0.15 | | | | |
| A | 395.74 | 5.60 | 0.26 | 391 | 2.36 | 0.13 | 387 | 391 |
| A | 405.78 | 7.23 | 0.33 | 422 | 6.96 | 0.36 | 422 | 424 |
| B | 411.42 | 6.35 | 0.28 | | | | |
| B | 428.70 | 5.74 | 0.25 | 432 | 10.06 | 0.50 | | |
| B | 627.57 | −6.90 | −0.20 | 713 | | | 728 | |
| A | 656.22 | −3.85 | −0.11 | | | | | |
| B | 661.86 | −9.36 | −0.26 | | | | | |
| A | 666.36 | −3.84 | −0.11 | | | | | |
| B | 673.73 | −6.06 | −0.17 | 734 | −2.93 | −0.09 | 733 | 735 |
| A | 681.87 | −5.62 | −0.15 | | | | | |
| A | 688.54 | −0.72 | −0.02 | 757 | −2.75 | −0.08 | 756 | 760 |
| A | 715.60 | −8.74 | −0.22 | 782 | −1.29 | −0.04 | 781 | 782 |
| B | 727.77 | −3.09 | −0.08 | | | | 796 | |
| A | 739.75 | −0.85 | −0.02 | | | | | |
| B | 755.66 | −2.08 | −0.05 | 817 | −0.22 | −0.01 | 815 | 815 |
| B | 767.03 | 2.02 | 0.05 | 840 | 7.79 | 0.20 | 838 | 836 |

# Figures

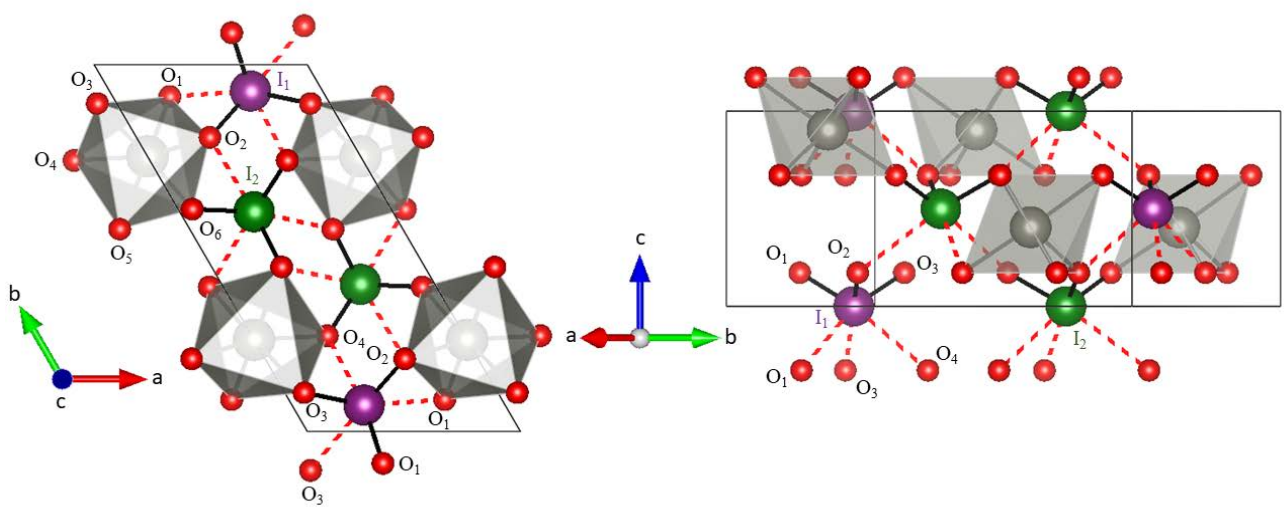

**FIG. 1.** Crystal structure of $Zn(IO_3)_2$ at ambient conditions. (a) perpendicular to the c axis and (b) along the c axis. I atoms are shown in purple ($I_1$) and green ($I_2$) color. The Zn coordination octahedra are shown in gray. O atoms are shown in red and have been labeled. First neighbor I-O bonds are shown with solid black lines and second neighbor I-O bonds are shown with red dashed lines.

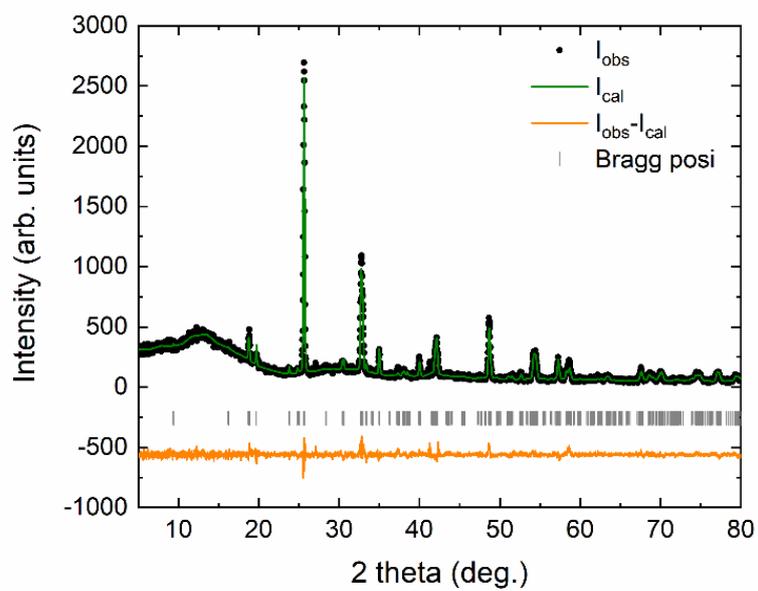

**FIG. 2.** XRD pattern measured at ambient pressure (circles). The green line is the Rietveld refinement. The quality factors of Rietveld refinement are $R_p$ = 7.22% and $R_{wp}$ = 9.91%. Residuals are shown in orange.

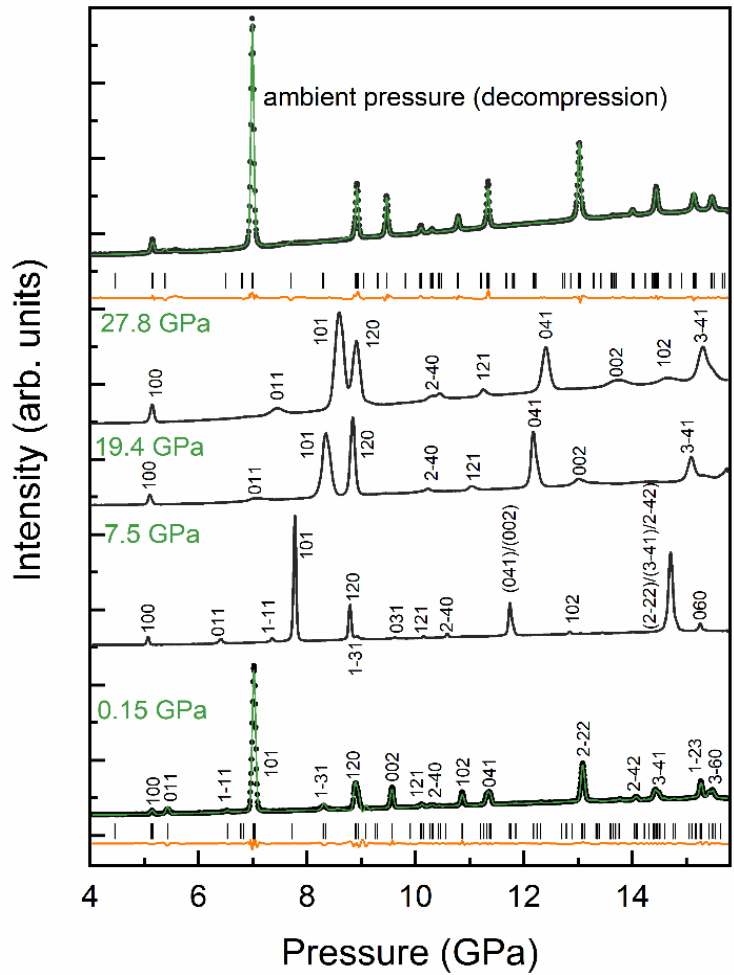

**FIG. 3.** Selected XRD patterns of $Zn(IO_3)_2$ under pressure. Bottom and top patterns show the refinement at the lowest pressures of the compression and decompression processes. Experimental data, refinement data and residuals are plotted with dots, green and orange lines, respectively. The Bragg peak positions (black vertical lines) are also shown.

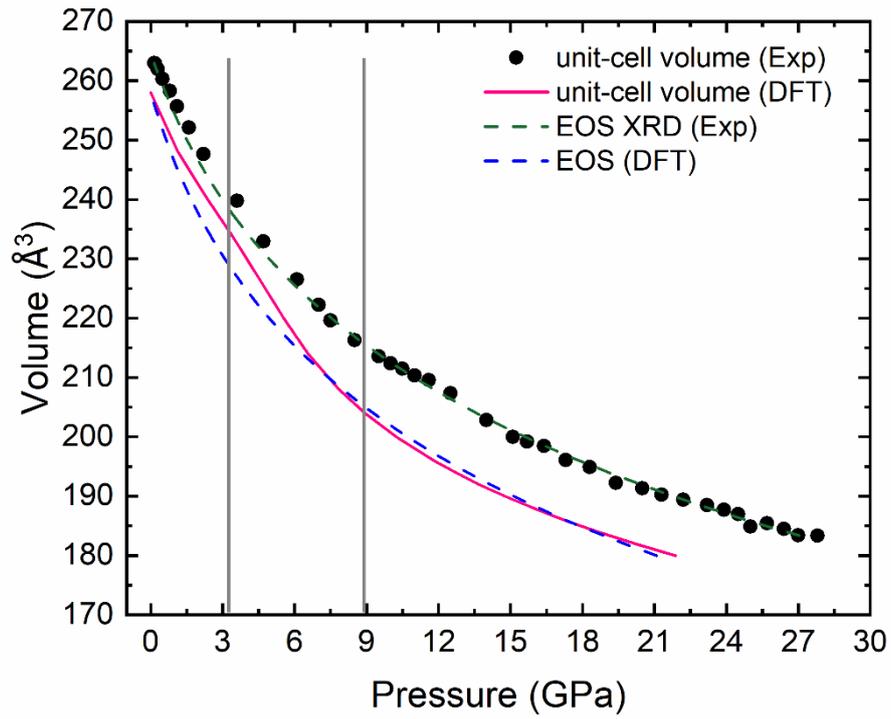

**FIG. 4.** Pressure dependence of the unit-cell volume. The black circles are the results of experiments and the magenta solid lines the results of calculations. The equations of state described in the text are shown with dashed lines. The vertical lines indicate the transition pressures.

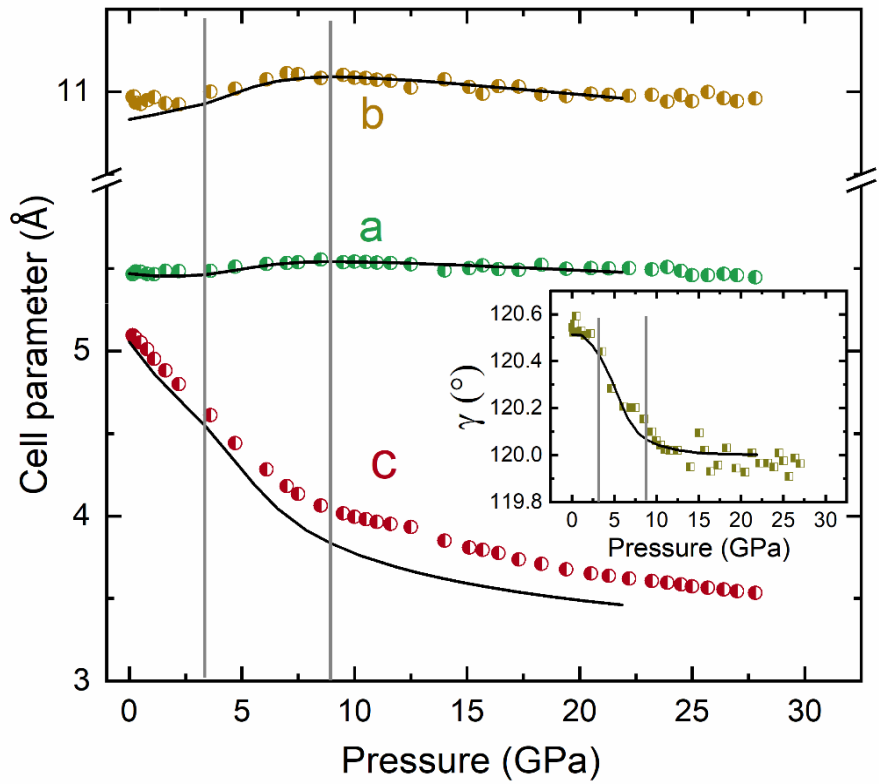

**FIG. 5.** Pressure dependence of the unit-cell parameters of $Zn(IO_3)_2$. The inset shows the evolution of the monoclinic angle $\gamma$. Half-filled circles or squares are the experimental data while DFT calculation results are shown in black solid lines.

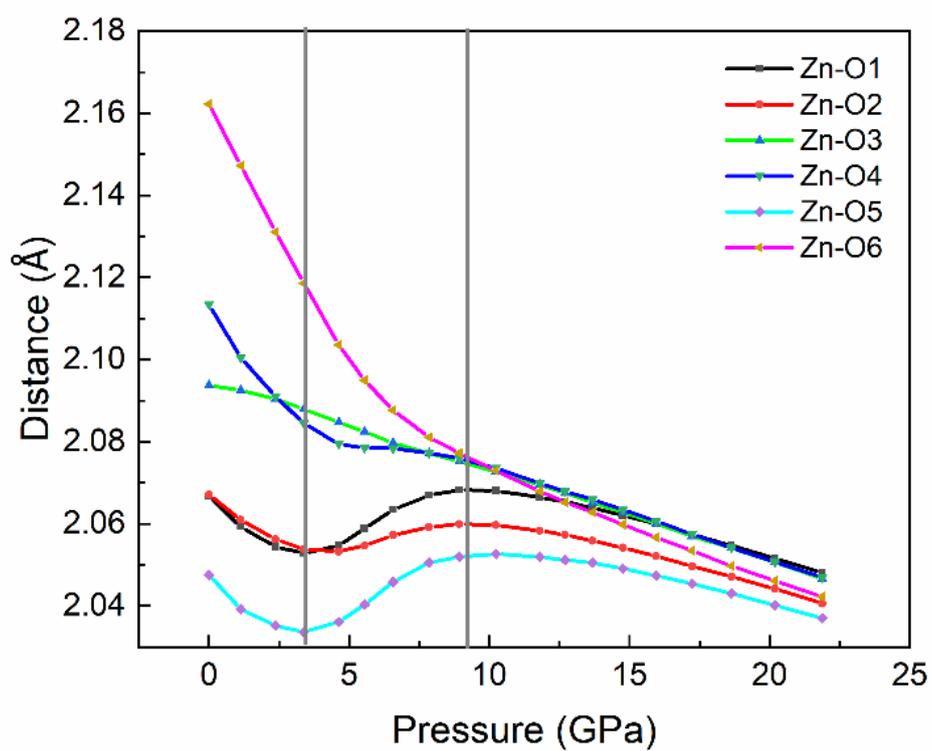

**FIG. 6.** Pressure dependence of the theoretical Zn-O bond distances.

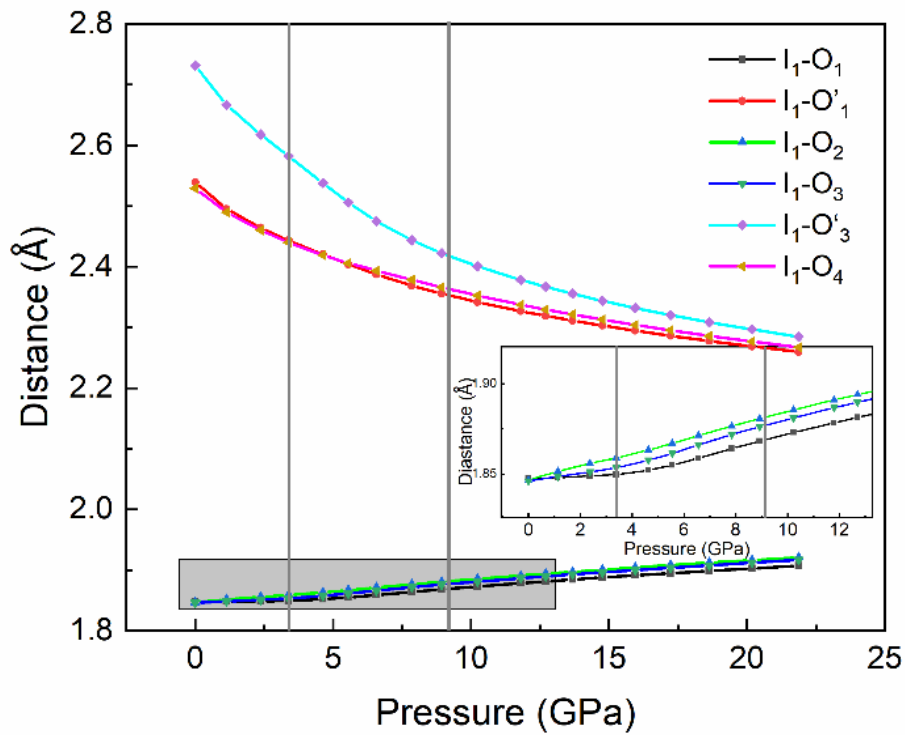

**FIG. 7.** Pressure dependence of the theoretical I-O bond distances. The inset shows a zoom of the I-O$_1$, I-O$_2$ and I-O$_3$ distances below 13 GPa.

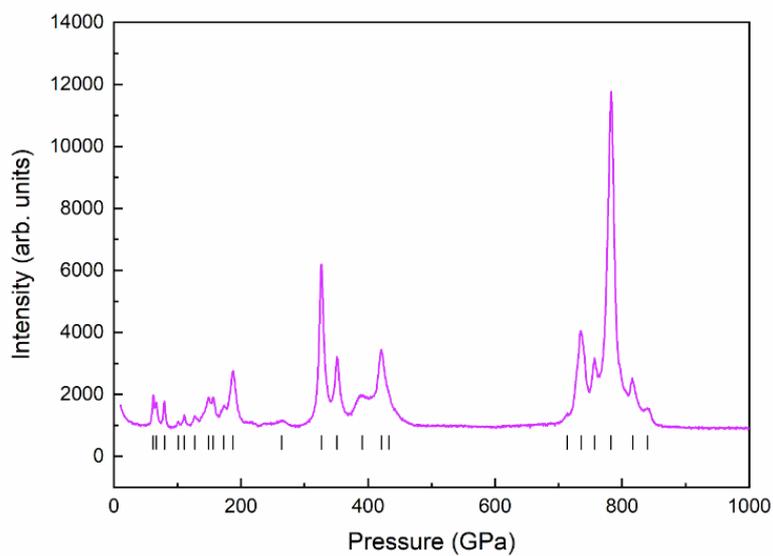

**FIG. 8.** Raman spectra of Zn(IO$_3$)$_2$ at ambient conditions. The determined peak position is indicated by ticks.

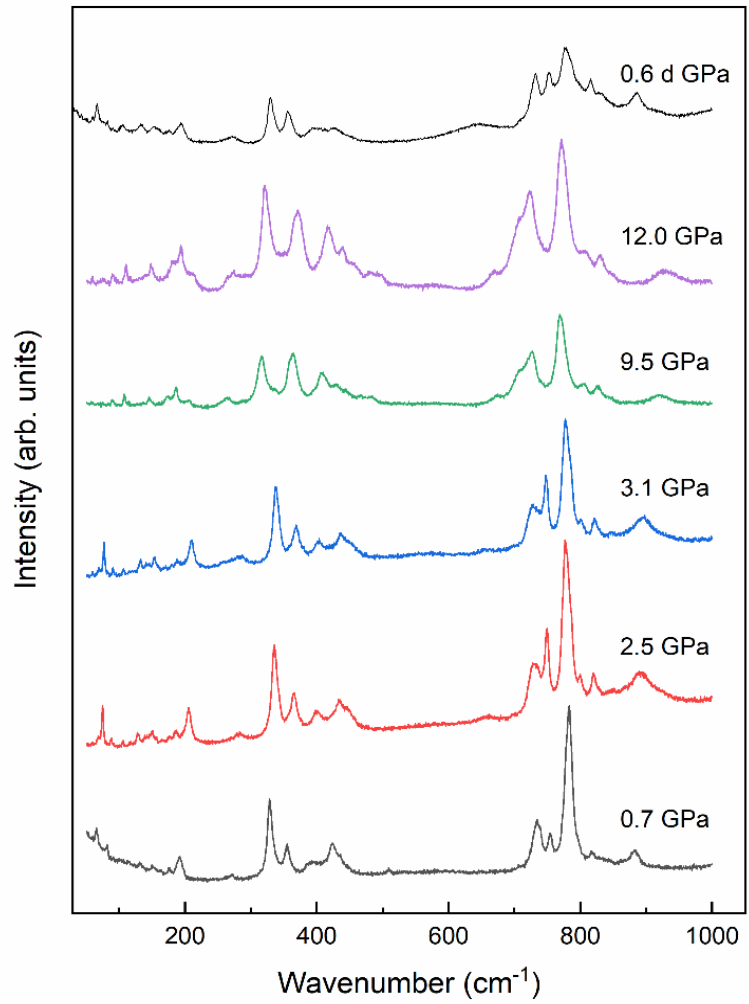

**FIG. 9.** Selected Raman spectra of $Zn(IO_3)_2$ under pressure. Pressure are indicated and "d" means decompression.

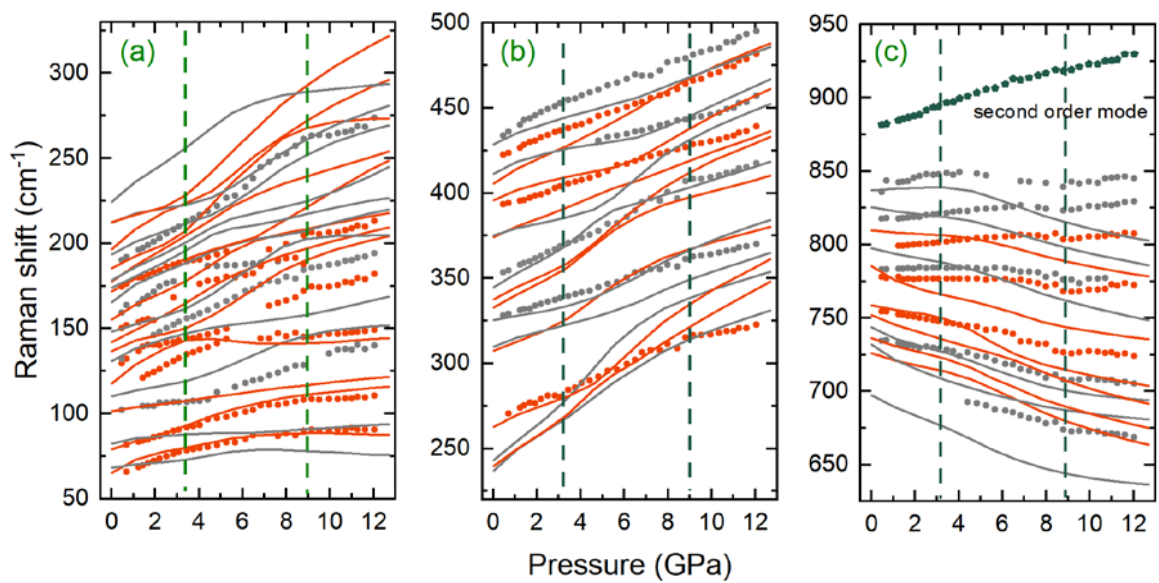

**FIG. 10.** Pressure dependence of experimental (symbols) and theoretical (lines) Raman-active modes of $Zn(IO_3)_2$. A tentative mode assignment of the experimental data is show with different colors: Red and gray colors are used for A and B modes, respectively. In (c), we shifted all calculated modes by +70 cm$^{-1}$, in order to facilitate the comparison with experiments.

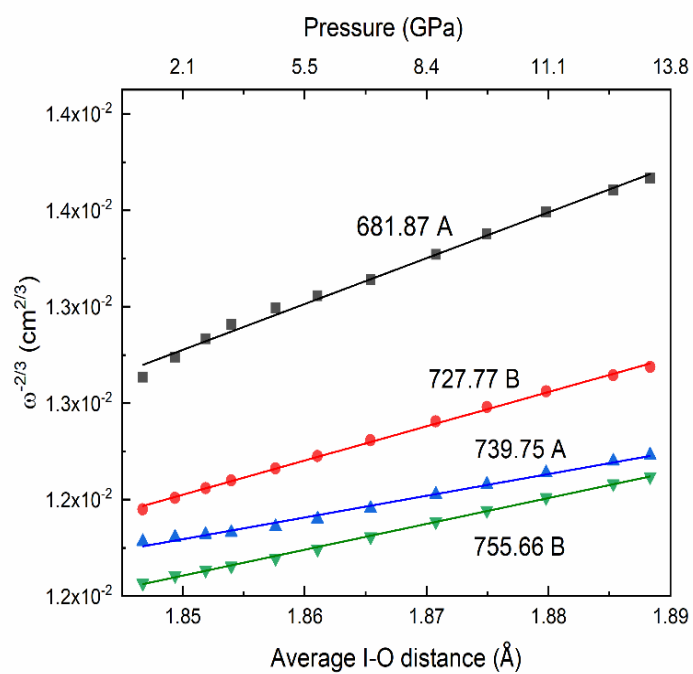

**FIG. 11.** Relationship between average I-O bonds and the theoretical $\omega^{-2/3}$ for some soft Raman-active modes of $Zn(IO_3)_2$. The corresponding atomic movement of each mode can be found in Fig. S13 [40]. In the upper axis the pressure of each average bond is indicated.